\documentclass[prl,twocolumn,amsmath,amsfonts,amssymb,showpacs,usenatbib]{revtex4} 
\usepackage{epsfig,psfrag,graphicx,bm,color,xspace}
\usepackage[normalem]{ulem}

\usepackage{natbib}

\newcommand{\Mstream}{{\it M-streaming}\xspace}
\newcommand{\Mflatturb}{{\it M-turbulence}\xspace}
\newcommand{\Mprimary}{{\it M-primaries}\xspace}

\newcommand{\rmn}{\mathrm}
\newcommand{\Vph}{V_\mathrm{ph}}
\newcommand{\mug}{\mu G}

\begin{document}

\title{Turbulence and Particle Acceleration in Giant Radio Halos: the Origin of Seed Electrons}  

\author{Anders Pinzke$^{1,2}$} \email{apinzke@fysik.su.se (AP)} 
\author{S. Peng Oh$^{3}$}\email{peng@physics.ucsb.edu (PO)}
\author{Christoph Pfrommer$^{4}$}\email{christoph.pfrommer@h-its.org (CP)}

\affiliation{$^{1}$The Oskar Klein Centre for Cosmoparticle Physics, Department
  of Physics, Stockholm University, AlbaNova University Center, SE - 106 91
  Stockholm, Sweden}

\affiliation{$^{2}$Dark Cosmology Center, University of Copenhagen,
  Juliane Maries Vej 30, DK-2100 Copenhagen, Denmark}

\affiliation{$^{3}$University of California - Santa Barbara,
  Department of Physics, CA 93106-9530, USA}

\affiliation{$^{4}$Heidelberg Institute for Theoretical Studies
  (HITS), Schloss-Wolfsbrunnenweg 35, 69118 Heidelberg, Germany}

\date{\today}


\begin{abstract}
About $1/3$ of X-ray-luminous clusters show smooth, unpolarized
radio emission on $\sim$Mpc scales, known as giant radio halos. One
promising model for radio halos is Fermi II acceleration of seed
relativistic electrons by turbulence of the intracluster medium (ICM);
Coulomb losses prohibit acceleration from the thermal pool. However,
the origin of seed electrons has never been fully explored. Here, we
integrate the Fokker-Planck equation of the cosmic ray (CR) electron
and proton distributions in a cosmological simulations of cluster
formation. For standard assumptions, structure formation shocks lead
to a seed electron population which produces too centrally
concentrated radio emission. Instead, we present three realistic
scenarios that each can reproduce the spatially flat radio emission
observed in the Coma cluster: (1) the ratio of injected turbulent
energy density to thermal energy density increase significantly with
radius, as seen in cosmological simulations. This generates a flat
radio profile even if the seed population of CRs is steep with
radius. (2) Self-confinement of energetic CR protons can be
inefficient, and CR protons may stream at the Alfv{\'e}n speed to the
cluster outskirts when the ICM is relatively quiescent. A spatially
flat CR proton distribution develops and produces the required
population of secondary seed electrons. (3) The CR proton to electron
acceleration efficiency $K_{ep} \sim 0.1$ is assumed to be larger than
in our Galaxy ($K_{ep} \sim 10^{-2}$), due to the magnetic geometry at
the shock. The resulting primary electron population dominates. Due to
their weaker density dependence compared to secondary electrons, these
primaries can also reproduce radio observations.  These competing
non-trivial solutions provide incisive probes of non thermal processes
in the high-$\beta$ ICM.
\end{abstract}

\maketitle

About one third of X-ray-luminous clusters show smooth, unpolarized
radio emission on $\sim$Mpc scales, known as giant radio halos (RHs)
\citep{2014IJMPD..2330007B}. They appear only in disturbed, merging
clusters and the RH luminosity correlates with the X-ray luminosity
\citep{2001A&A...369..441G,2012A&ARv..20...54F} and the Compton
$y$-parameter \citep{2013A&A...554A.140P}. The RHs show that CRs and
magnetic fields permeate a large volume fraction of the intra-cluster
medium (ICM). The dominant CR source, given the smoothness and
enormous extent of RHs, is thought to be structure formation shocks
\citep{miniati01,pfrommer08}. At the same time, plasma processes, the
origin of magnetic fields and particle acceleration in a turbulent,
high-$\beta$ plasma like the ICM are not well understood. Radio halos
thus provide an incisive probe of non-thermal processes in the ICM.

One promising model for RHs is re-energetization of seed suprathermal
electrons by Fermi II acceleration when ICM turbulence becomes transonic during
mergers \citep{2001MNRAS.320..365B,brunetti07}. Due to the short radiative
cooling time of high-energy relativistic electrons, the cluster synchrotron
emission quickly fades away after a merger, which naturally explains the
observed bimodality of RHs.

\begin{figure*}
  \begin{minipage}{\columnwidth}
    \large{\Mflatturb:}\\ 
    \includegraphics[width=\columnwidth]{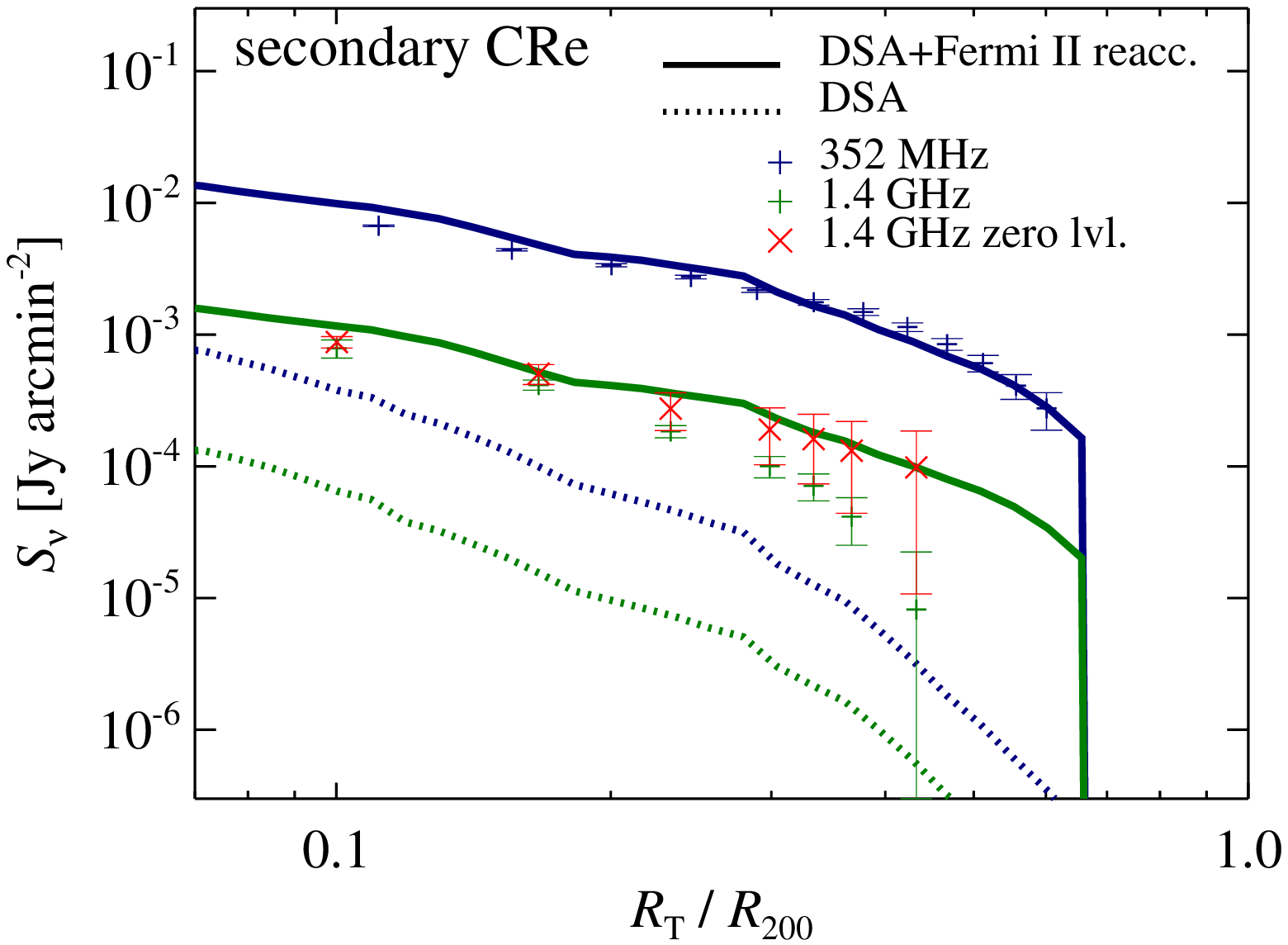}
  \end{minipage}
  \begin{minipage}{\columnwidth}
    \large{\Mstream:}\\
    \includegraphics[width=\columnwidth]{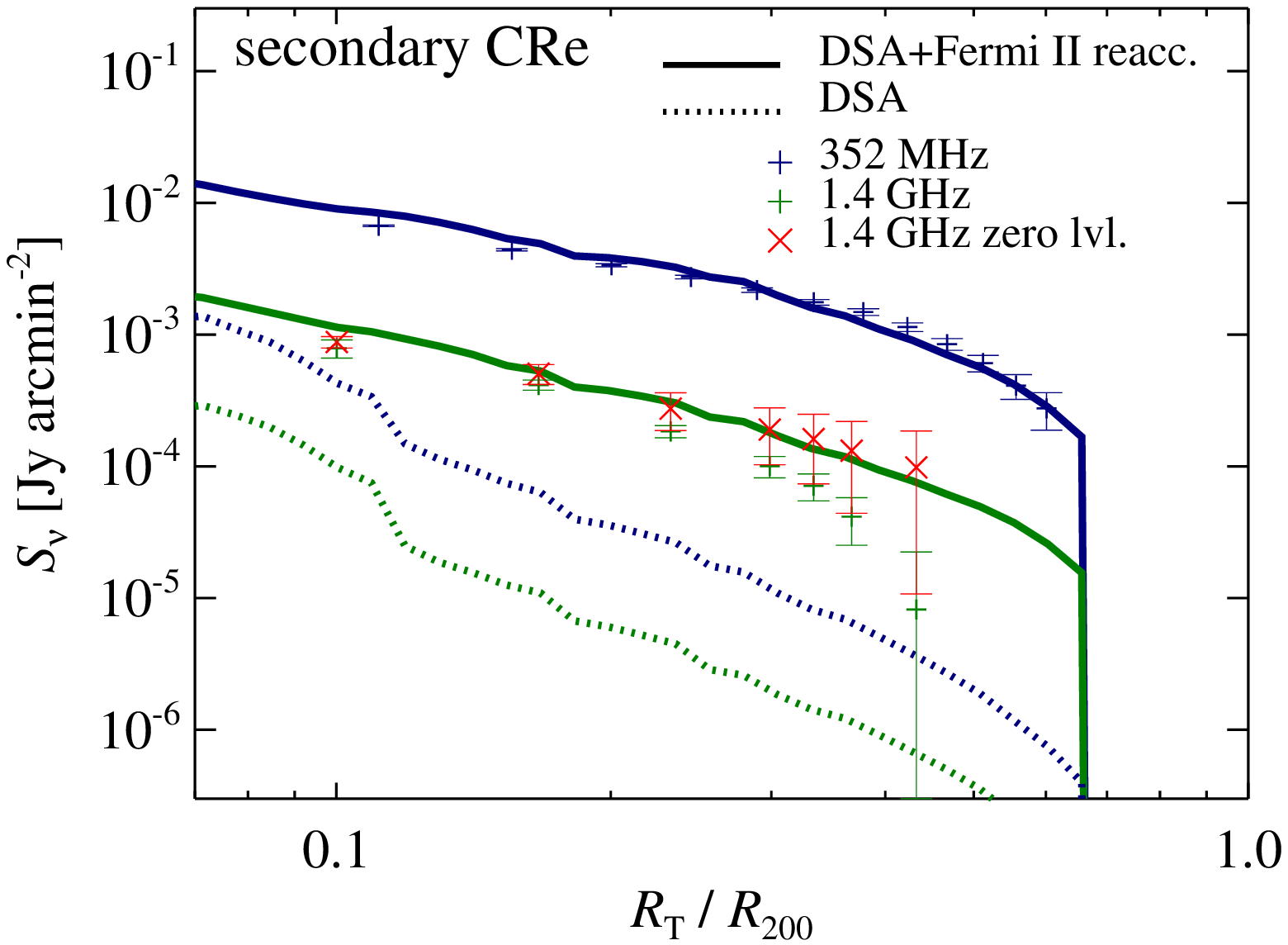}
  \end{minipage}
  \\
  \begin{minipage}{\columnwidth}
    \large{\Mprimary:}\\
    \includegraphics[width=\columnwidth]{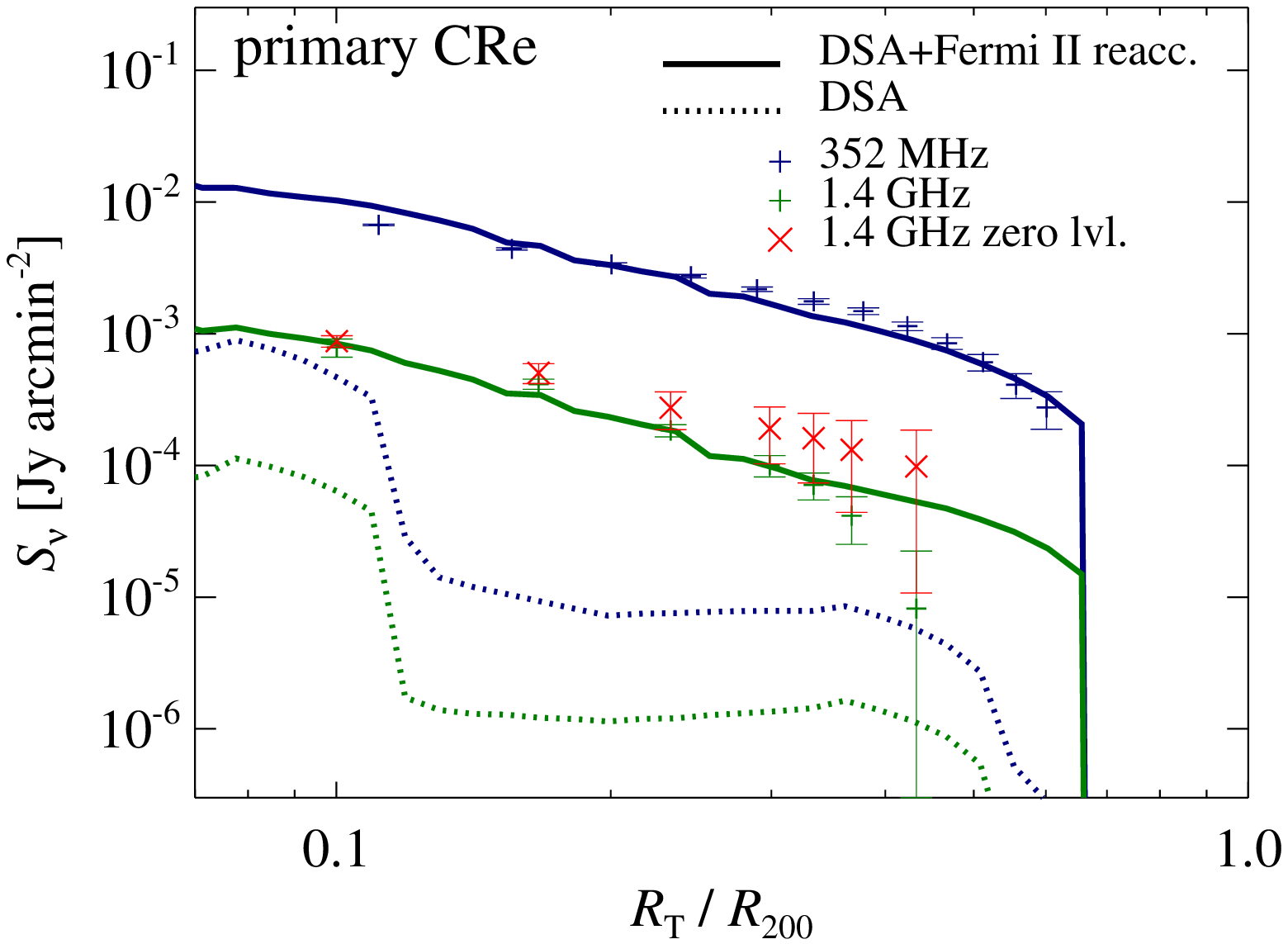}
  \end{minipage}
  \begin{minipage}{\columnwidth}
    \large{\it Brunetti et al. (2012)}:\\
    \includegraphics[width=\columnwidth]{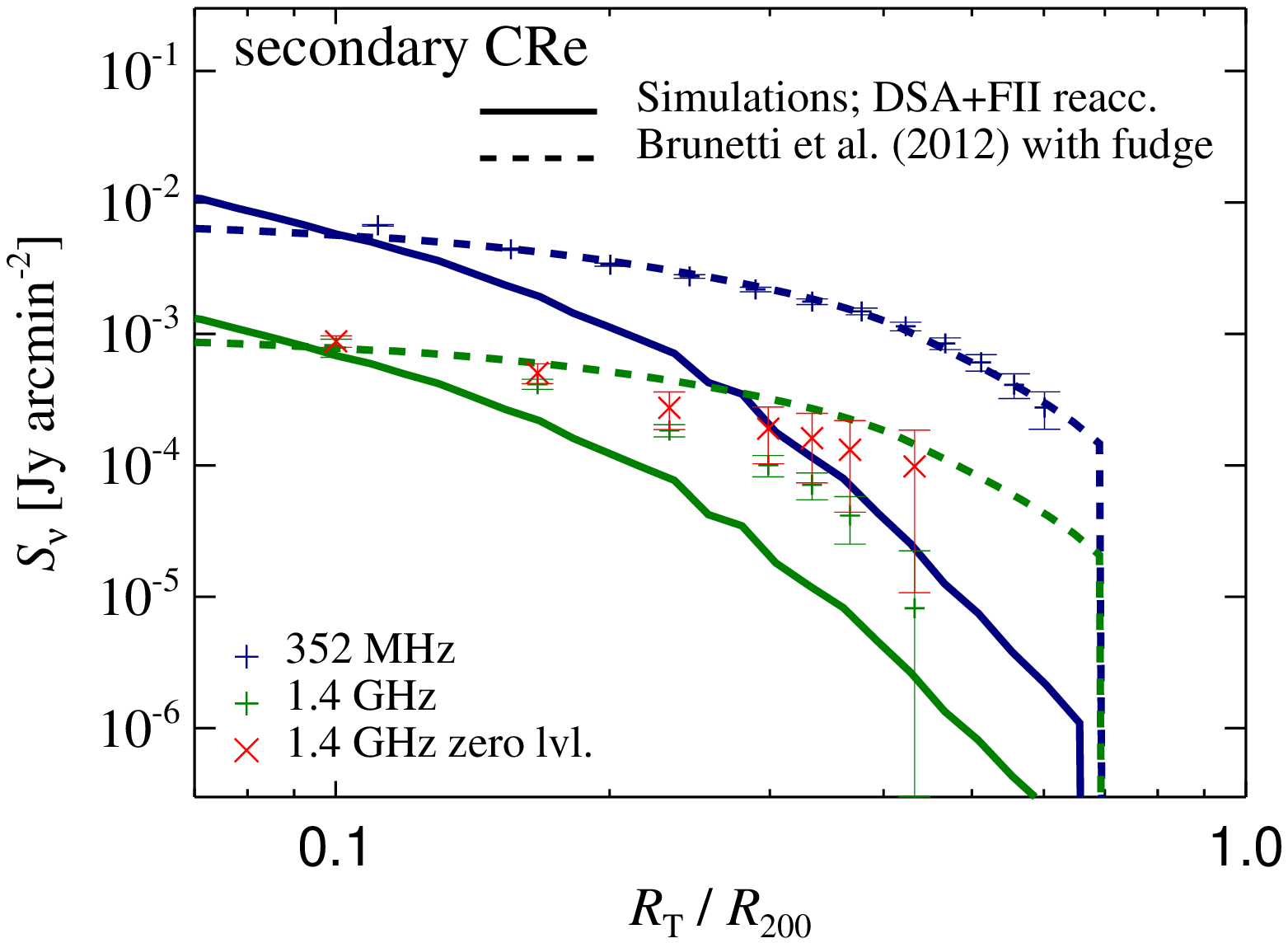}
  \end{minipage}
  \caption{Radio surface brightness profiles of Fermi-II reaccelerated
    CR electrons of a simulated post-merging cluster similar to
    Coma. We compare profiles at 352~MHz (blue lines and crosses
    \cite{brown11}) to those at 1.4~GHz (green lines and crosses
    \cite{deiss97}). The red crosses show the reprocessed 1.4~GHz
    data, where a zero level of about 10 \% of the central value is
    adopted. The solid lines show predicted emission from a
    reaccelerated fossil population, while dotted lines show emission
    from a fossil population without reacceleration. The panels show
    the emission from CR protons and secondary electrons reaccelerated
    by a flat turbulent profile (upper left panel), secondary
    electrons generated by streamed CR protons (upper right panel),
    primary electrons (lower left panel), and simulated secondary
    electrons together with previous estimates \cite{brunetti12} for
    the Coma cluster (lower right panel).}
 \label{fig:sync_profile}
\end{figure*}

However, there is a salient piece missing in the turbulent
reacceleration model. It relies heavily on the assumption of an
abundant, volume-filling population of seed suprathermal electrons;
direct Fermi II acceleration from the thermal pool is precluded by
strong Coulomb losses
\citep{2008ApJ...682..175P,2012ApJ...759..113C}. These seeds are
presumed to be either fossil CR electrons (CRes) accelerated by
diffusive shock acceleration (DSA) during structure formation
\citep{1999ApJ...520..529S}, or secondaries injected by hadronic
interaction of CR protons (CRps) with thermal protons
\citep{brunetti11}. While analytic estimates have been made, there has
been no ab initio demonstration that structure formation can lead to
the required abundance of seed electrons with the correct spatial and
spectral characteristics. This is a non-trivial requirement: Coulomb
cooling in dense cluster cores is severe, and DSA fossil electrons may
not survive. On the other hand, for secondaries to constitute the seed
population, the CRp population required in the best-studied case of
the Coma cluster must have a very broad and flat (or even slightly
inverted) spatial profile \citep{brunetti12}, in contrast with the
thermal plasma whose energy density declines steeply with radius. In
this Letter we show that such a distribution is not predicted by
cosmological simulations (see lower right panel of
Fig.~\ref{fig:sync_profile}) \citep[see
  also][]{pinzke10,2014MNRAS.439.2662V}.

Indeed, arriving at a seed population with the required
characteristics is highly constraining, and has the potential to teach
us much about the origin of CRps/CRes in clusters. We consider 3 new
possibilities: (i) Our model \Mflatturb: a significantly flatter
turbulent profile than what was adopted in \cite{brunetti12}, which
allows seed CRps to follow the steep profile that is suggested by
structure formation simulations. (ii) Our model \Mstream: streaming
CRps that produce flat distributions of CRs in the ICM
\citep{ensslin11,wiener13}, which also flattens the secondary electron
distribution. (iii) Our model \Mprimary: if the acceleration
efficiency of CRps is below about $0.1$~{\%} in weak (perpendicular)
shocks and the ratio of injected electrons-to-protons $K_{\rmn{ep}}
\sim 0.1$, this yields a dominant primary population with a flat
spatial distribution, since primaries have a weaker density dependence
than secondaries. In this work we pursue these three possibilities
further. We employ cosmological simulations of CRs in clusters, in
tandem with new insights from our recent work on DSA generated fossil
electrons \citep{pinzke13}, to generate the first quantitative
calculation of primary and secondary seed electrons.

{\bf Method.}  The transport of relativistic electrons and protons in
the ICM is a complex process that depends both on the details of the
thermal component (gas density, temperature, and pressure) as well as
non-thermal component (turbulence, magnetic fields, fossil CRs). We
use high resolution galaxy cluster simulations to derive the thermal
and fossil CR properties (shock accelerated primary CRes and CRps, as
well as secondary CRes produced in p-p collisions)
\cite{2007MNRAS.378..385P,pfrommer08,pinzke10,pinzke13}. In this
Letter we focus on our simulated cluster, g72a, which is a massive
$1.6\times10^{15}\,M_\odot$ cluster that experienced a merger about
1-2 Gyrs ago. Since the cluster mass, density and temperature profiles
are all similar to the well studied Coma cluster
\cite{2007MNRAS.378..385P,pinzke10}, we will compare our calculations
to radio and gamma-ray observations of Coma.

In our Galaxy, the CRe-to-CRp ratio at a few GeV is $K_{\rmn{ep}} \sim
10^{-2}$. Hence, we adopt this as a fiducial value for the CRe-to-CRp
acceleration efficiency (see \cite{pinzke13} for more
discussion). However, as recent PIC simulations have shown, this is
likely very different at weak shocks, with electrons efficiently
accelerated at perpendicular shocks
\citep{2014ApJ...794..153G,2014ApJ...797...47G}, and ions efficiently
accelerated at parallel shocks \citep{2014ApJ...783...91C}. Thus,
depending on magnetic geometry, $K_{\rmn{ep}}$ could be either larger
or smaller. In this work we use a simple test-particle model for the
CRp acceleration \cite{kang11,pinzke13}. The ratio of accelerated
proton-to-dissipated energy in the downstream of strong shocks varies
from 1-10\%, depending on the adopted model (for more details, see the
Results section), and is a factor 10-100 lower for weak
shocks. However, some observations of radio relics suggest higher
values of $K_{\rmn{ep}}$, due to the absence of gamma-ray emission,
which probes the CRp population \cite{2014MNRAS.437.2291V}.  This
suggests primary CRes as a viable alternative scenario to secondary
CRes as seeds for the giant RHs. In our {\it M-primaries} scenario, we
adopt $K_{\rmn{ep}} =0.1$ (viable for primarily perpendicular shocks)
to test this possibility.

As previously noted, secondaries produced by shock accelerated CRp
have the wrong spatial profile to explain RH observations; because
they arise from a two body process, they are too centrally
concentrated. They also produce $\gamma$-ray emission in excess of
Fermi-LAT upper limits
\citep{2012ApJ...757..123A,brunetti12,2014ApJ...787...18A}. However,
if CRps stream in the ICM, then their spatial profile could
potentially flatten sufficiently \citep{ensslin11,wiener13}. This
scenario is very attractive: it generates seed electrons with the
right spatial footprint, and by removing CRps from the core, obeys
gamma-ray constraints. Turbulence plays two opposing roles:
Alfv{\'e}nic turbulence damps waves generated by the CR streaming
instability \citep{yan02,farmer04}, thus reducing self-confinement;
but compressible fast modes scatter CRs directly. Turbulent damping is
still efficient for highly subsonic conditions \citep{wiener13}, while
we assume compressible fast modes only provide effective spatial
confinement during the periods of transonic, highly super-Alfv{\'e}nic
($M_{\rm A} \sim 5$) turbulence associated with mergers. Thus, CRs can
stream out when the cluster is kinematically quiescent. Furthermore,
even Alfv{\'e}nic streaming timescales are relatively short ($\sim
0.1-0.5$ Gyr; \cite{wiener13}) compared to the timescale on which the
CRp population is built up. Based on these findings, we adopt a toy
model for our \Mstream scenario in which CR streaming quickly produces
flat CRp profiles. We assume that CRs cannot
stream significantly past perpendicular $B$-fields at the accretion
shock, so that the total number of CRs is conserved.

Given a seed population of CRs, we adopt essentially the same set of
plasma physics assumptions as the reacceleration model for RHs
\cite{brunetti07,brunetti11}, leaving exploration of parameter space
to future work. We solve the isotropic, gyro-phase averaged
Fokker-Planck equation (via a Crank-Nicholson scheme) for the time
evolution of the CRe distribution in the Lagrangian frame
\cite{brunetti07,brunetti11}:
\begin{eqnarray}
{{d f_{\rmn{e}}(p,t)}\over{d t}} &\!=&
\frac{\partial}{\partial p}
\left\{
f_{\rmn{e}}(p,t)\left[
\left|{{dp}\over{dt}}\right|_{\rm C} 
+ \frac{p}{3}\left(\vec{\nabla}\cdot \vec{\upsilon}\right)\right.\right.
\nonumber\\
&+& \left.\left. \left|{{dp}\over{dt}}\right|_{\rm r}
- {1\over{p^2}}{{\partial }\over{\partial p}}\left(p^2 D_{pp}\right) 
\right]\right\} - \left(\vec{\nabla}\cdot \vec{\upsilon}\right) f_{\rmn{e}}(p,t)
\nonumber\\
&+& {{\partial^2 }\over{\partial p^2}}
\left[
D_{pp} f_{\rmn{e}}(p,t) \right]+ Q_{\rmn{e}}\left[p,t;f_{\rmn{p}}(p,t)\right]   \, .
\label{elettroni}
\end{eqnarray}
Here $f_{\rmn{e}}$ is the one-dimensional distribution in position $x$
(suppressed for clarity), momentum $p$ and time $t$ (which is
normalized such that the number density is given by
$n_{\rmn{e}}(t)=\int d p f_{\rmn{e}}(p,t)$), $d/dt=\partial/\partial
t+\vec{\upsilon}\cdot \vec{\nabla}$ is the Lagrangian derivative,
$\vec{\upsilon}$ is the gas velocity, $|dp/dt|$ represents radiative
(r) and Coulomb (C) losses, $D_{pp}$ is the momentum space diffusion
coefficient, and $Q_{\rmn{e}}$ denotes the injection rate of primary
and secondary electrons in the ICM. The $\vec{\nabla}\cdot
\vec{\upsilon}$ terms represent adiabatic gains and losses. During
post-processing of our Coma-like cluster simulation, we solve the
Fokker-Planck equation over a redshift interval from $z=5$ to 0. The
simulated cluster undergoes a major merger over the last 1-2~Gyrs that
injects large turbulent eddies. After about 1~Gyr those have decayed
down to the scale needed to reaccelerate particles. In all our
calculations we assume that turbulent reacceleration is efficiently
accelerating particles for 650 Myrs and that during this turbulent
phase CR streaming and spatial diffusion can be neglected. In our
\Mstream model, CR streaming and diffusion are incorporated separately
during kinematically quiescent times that precede the merger. As a
result, flat CRp profiles are produced on relatively short timescales
($\sim 0.1-0.5$ Gyr).

The time evolution of the spectral energy distribution of CRps,
$f_{\rmn{p}}$, is similarly given by:
\begin{align}
{{d f_{\rmn{p}}(p,t)}\over{d t}} &=
{{\partial }\over{\partial p}}
\left\{
f_{\rmn{p}}(p,t)\left[ \left|{{dp}\over{dt}}\right|_{\rm C}
+ \frac{p}{3}\left(\vec{\nabla}\cdot \vec{\upsilon}\right)\right.\right.
\nonumber\\
&-\left.\left. {1\over{p^2}}{{\partial }\over{\partial p}}\left(p^2 D_{pp}\right)
\right]\right\} - \left(\vec{\nabla}\cdot \vec{\upsilon}\right) f_{\rmn{p}}(p,t)
\nonumber\\
&+ {{\partial^2 }\over{\partial p^2}}
\left[ D_{pp} f_{\rmn{p}}(p,t) \right] - {{f_{\rmn{p}}(p,t)}\over{\tau_{\rm had}(p)}}
+ Q_{\rmn{p}}(p,t)\, ,
\label{eq:FP_p}
\end{align}
where $Q_{\rmn{p}}(p,t)$ denotes the injection rate of shock
accelerated CRps as a function of momentum $p$ and time $t$, and
$\tau_{\rm had}$ is the timescale of hadronic losses that produce
pions via CRp collisions with thermal protons of the ICM
\cite[e.g.][]{brunetti11}.  We incorporate momentum diffusion for
electrons and protons from transit-time-damping (TTD) resonance with
compressible magneto-hydrodynamic (MHD) turbulence, to model Fermi-II
reacceleration \cite{brunetti07,brunetti11}. The TTD resonance
requires the wave frequency $\omega=k_\parallel\upsilon_\parallel$,
where $k_\parallel$ and $\upsilon_\parallel$ are the parallel
(projected along the magnetic field) wavenumber and particle velocity,
respectively. This implies that the particle transit time across the
confining wave region matches the wave period,
$\lambda_{\parallel}/v_{\parallel}=T$. The resonance changes the
component of particle momentum parallel to seed magnetic fields, which
over time leads to increasing anisotropy in the particle distribution
that decreases the efficiency of reacceleration with time. As in
ref.{\ }\cite{brunetti11}, we assume that there exists a
mechanism---such as the firehose instability---that isotropizes the CR
distribution function at the gyroscale and on the reacceleration time
scale, which ensures sustained efficient reacceleration with time. The
particle pitch-angle averaged momentum-diffusion coefficient of
isotropic particles that couple to fast magnetosonic modes via TTD
resonance is \cite{brunetti07} (Eqn. 47):
\begin{eqnarray}
  D_{pp}(p,t) = \frac{\pi}{16} \frac{p^2}{c\,\rho}
  \left\langle\frac{\beta |B_k|^2}{16 \pi \,W}\right\rangle
  I_\theta
  \int_{k_\rmn{cut}}\mathcal{W}(k)k\,d k\,,
\label{eq:dpp}
\end{eqnarray}
where $\beta$ is the thermal-to-magnetic pressure ratio, and $c$
is the speed of light. The energy density $W$ of a mode in a
magnetized plasma stems from both electromagnetic fields and resonant
particles. For a high-$\beta$ plasma, the pitch angle averaged ratio of
beta-weighted magnetic-to-total energy density saturates to $\langle\beta
|B_k|^2/2W\rangle\approx 10^{1.4}$ (see figure 2 in
\cite{brunetti07}). The pitch angle of the CR momentum with the
magnetic field orientation is given by $\theta$, and
$I_\theta=\int_0^{\arccos(\Vph/c)} d\theta {\frac{ \sin^3 \theta }{
    |\cos \theta | }}
\left[1+\left(\frac{\Vph}{c\,\cos{\theta}}\right)^2\right]^2$. Here
$V_\rmn{ph}$ is the phase velocity of the fast magnetosonic waves
given approximately by the sound speed, $\Vph \sim c_\rmn{s}$. For a sound
speed typical for the ICM of 1000~km/s, $I_\theta\approx5$. As in
\cite{brunetti07}, we initially assume that the velocity of turbulent
eddies is $V_0\approx 0.47 c_\rmn{s}$ throughout the cluster. This gives a
turbulent acceleration time scale, $\tau_{pp} = p^2/4D_{pp}$, that is
typically few 100 Myrs in the ICM. 

We adopt a simplified isotropic Kraichnan MHD turbulent spectrum for
the fast modes per elemental range $dk$ of the form
\begin{equation}
  \mathcal{W}(k) \approx \sqrt{I_0\,\rho\,\langle \Vph \rangle}\,k^{-3/2}\,,
\end{equation}
for $k_0<k<k_\rmn{cut}$, where we assume an injection scale for the
turbulence, $k_0= 2\pi/(100~\mbox{kpc})$. The volumetric injection
rate of turbulent energy, $I_0$, is fixed by requiring that the total
turbulent energy density on the largest scales
$\epsilon_\rmn{turb}=\int \mathcal{W}(k) dk \approx 0.2
\epsilon_\rmn{th}$, where $\epsilon_\rmn{th}$ is the thermal energy
density \cite{brunetti07,brunetti11}. In this work we investigate
different spatial models for injected turbulence. We assume that
$\epsilon_\rmn{turb} \propto \epsilon_\rmn{th}^{\alpha_\rmn{tu}}$,
where $\alpha_\rmn{tu}= 0.69$ for \Mflatturb, $\alpha_\rmn{tu}= 0.84$
for \Mstream, and $\alpha_\rmn{tu}= 0.91$ for \Mprimary (note that in
previous work, $\alpha_\rmn{tu}= 1$ was adopted
\cite{brunetti12}). Our flatter turbulent profiles are motivated by
fits to cosmological simulations
\cite{2009ApJ...705.1129L,2010ApJ...725.1452S,2012ApJ...758...74B} and
the range indicates uncertainties of the turbulent profile in
Coma. Future observations (by Astro-H) and simulations will help to
clarify this issue. Provided dissipation of turbulence in the ICM is
collisionless, turbulent cascades of compressible modes become
suppressed when thermal and relativistic particles resonantly interact
with magnetosonic waves via TTD on a timescale $\Gamma^{-1}$ that is
approaching the cascading timescale given by $\tau_{kk} \approx
k^2/D_{kk}$. Here the wave-wave diffusion coefficient of magnetosonic
modes is given by
\begin{equation}
  D_{kk} \approx \Vph k^4 \left(\frac{\mathcal{W}(k)}{\rho\,\Vph^2}\right)\,.
\end{equation}
Thus, the cascade is suppressed for wave numbers above:
\begin{equation}
\label{eq:kc}
  k_\rmn{cut} \approx \frac{81}{14} \frac{I_0}{\rho \langle \Vph \rangle}
  \left(\frac{\langle\sum_i \Gamma_i(k,\theta)\rangle}{k}\right)^{-2}\,.
\end{equation}
where $2\pi/k_\rmn{cut}\sim 0.1-1$~kpc in the ICM. This constitutes an
effective mean free path for CRs, unless plasma instabilities can
mediate interactions between turbulence and particles on smaller
scales \cite{brunetti11}. In this work we only consider damping via
TTD due to thermal electrons, and neglect subdominant damping with
thermal protons and relativistic particles. The latter will be
subdominant in the ICM for a CR to thermal energy density ratio
$\lesssim 10 \%$ \cite{brunetti07}, which is always satisfied. The
azimuthally averaged turbulent damping rate from thermal electrons
\cite{brunetti07} in a high-$\beta$ plasma is $
\langle\Gamma_{\rmn{e}}\rangle \approx \langle k\,\Vph\,\sqrt{3\pi
  x/20}\exp(-5x/3)\sin^2{\theta}\rangle\approx 0.0435k\,\Vph$, where
$x=(m_{\rmn{e}}/m_{\rmn{p}})/\cos^2{\theta}$. The magnetic field in
the ICM is typically $\sim \mug$. To compute the synchrotron surface
brightness profiles, we use the profile of the magnetic field strength
derived from Faraday rotation observations of Coma \cite{bonafede10}
in combination with the density profile derived from X-ray
measurements \cite{1992A&A...259L..31B}.

{\bf Results and Discussion.}  Let us first consider the two models
which rely on secondary electrons. After turbulent reacceleration, the
volume-weighted, relative CRp energy density and CRp number density
inside the RH for \Mflatturb (\Mstream), are found to be 3
(2) \% and $3.0\times10^{-8}$ ($4.5\times10^{-8}$), respectively. As
we will see later, these densities are just of the right order of
magnitude to reproduce radio observations in the Coma cluster. In
addition we predict the gamma-ray flux within the virial radius of the
Coma cluster from CRps that produce decaying neutral pions for
\Mflatturb (\Mstream) with
$F_\gamma(>500\,\rmn{MeV})=1.6\times10^{-10} (2.3\times10^{-10})
\mathrm{ \,ph\, s}^{-1}\mathrm{cm}^{-2}$. Both fluxes are well below
current limits set by Fermi-LAT \cite{2014ApJ...787...18A}, and will
be challenging to probe in the near future. The spectral index of the
CRp distribution is relatively steep $\alpha_{\rmn{p}}\sim2.6$ for the
CRp energies $E \gtrsim 10$~GeV that are relevant for the injection of
radio-emitting secondary CRes.  The steep spectrum is ultimately a
consequence of our test particle model for Fermi-I acceleration
\cite{pinzke13}, where we steepen the spectral index to avoid
acceleration efficiencies above $\zeta_{\rmn{p}}=15\%$.

In Fig.~\ref{fig:sync_profile}, we find that all three scenarios in
which the seeds undergo Fermi-II reacceleration can reproduce the Coma
RH profile at 352~MHz. In the panel {\it Brunetti et al. (2012)} we
show that without CR streaming or a flat turbulent profile, our
simulations of reaccelerated CRs produce radio profiles that are too
steep. Indeed, even using the assumptions of previous work -- where
complete freedom in the seed population was allowed -- it is not
possible to reproduce observations in both frequencies. This signals
that this problem is generic and requires either additional
modifications to the plasma physics of Fermi-II acceleration or a
better understanding of potential observational systematics. 

\begin{figure}
  \includegraphics[width=0.99\columnwidth]{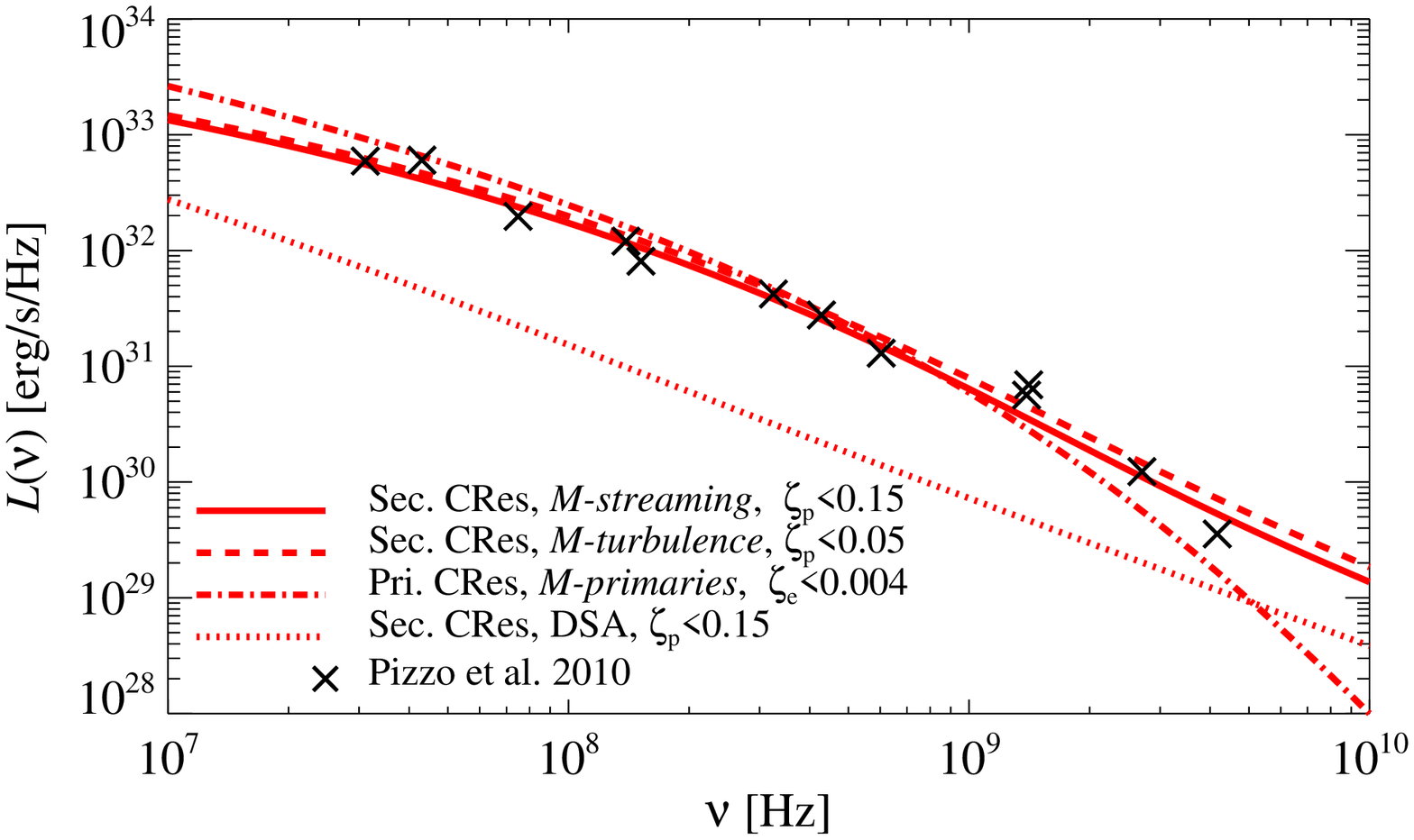}
  \caption{Radio synchrotron flux as a function of frequency. Red
    lines are derived from simulations, while the black crosses are
    compiled from observations \cite{2010PhDT.......259P}. The lines
    show the emission from secondary electrons generated by streamed
    CR protons (solid line), CR protons and secondary electrons
    reaccelerated by a flat turbulent profile (dashed line), and
    primary electrons (dash-dotted line). We contrast the
    reaccelerated populations to a population of DSA accelerated CR
    protons that produce secondaries (dotted line).}
 \label{fig:sync_spectrum}
\end{figure}

In principle, reacceleration via TTD leads to spectral steepening with
particle energy due to the inefficiency of the acceleration process to
counter the stronger cooling losses with increasing energy. Since
synchrotron emission peaks at frequency $\nu_\rmn{syn}\simeq 1\,
B/\mu\rmn{G} (\gamma/10^4)^2\,\rmn{GHz}$, this translates into a
spectral steepening of the total radio spectrum (see
Fig.~\ref{fig:sync_spectrum}). A given radio window samples higher
energy electrons for a decreasing field strength in the cluster
outskirts. Hence, the spectral steepening with energy should translate
into a radial spectral steepening \cite{brunetti12}. However, because
of the weak dependence of the electron Lorentz factor on emission
frequency ($\gamma\propto\sqrt{\nu_\rmn{syn}}$), this effect is only
visible in our simulations for $\nu_\rmn{syn}\gtrsim5$~GHz. Most
importantly, our simulated fluid elements at a given radius sample a
broad distribution of shock history, density and temperature, which
implies very similar synchrotron brightness profiles at
$\nu_\rmn{syn}=352$~MHz and 1.4 GHz. The discrepancy of the observed
and simulated 1.4 GHz profiles could instead be due to systematic flux
calibration error in single dish observations. Interestingly, we can
match the 1.4 GHz data if we reduce the zero point by adding 10\% of
the central flux to every data point; this flattens the outer
profile \footnote{Lawrence Rudnick, private communication.}.
Alternatively, this may point to weaknesses in the theoretical
modeling of the particle acceleration process and may require a
stronger cutoff in the particle energy spectrum.
  
In Fig.~\ref{fig:sync_spectrum} we show that our three models that
include Fermi-II reacceleration can individually reproduce the
convexly curved total radio spectrum found in the Coma cluster. Seed
CRs that do not experience turbulent reacceleration have a power-law
spectrum in disagreement with observations. In order to match both the
spatial and spectral profiles in Coma, we can constrain the
acceleration efficiency for the strongest shocks in our three models
\Mstream, \Mflatturb, and \Mprimary to $\zeta_{\rmn{p}} < 0.15$,
$\zeta_{\rmn{p}}<0.05$, and $\zeta_{\rmn{e}} <0.004$,
respectively. Following the Mach number ($\mathcal{M}$)-dependence of
the acceleration efficiency suggested in \cite{pinzke13}, the
efficiency in weak shocks ($\mathcal{M}\sim 2.5-3.5$) that dominates
the CR distribution function, has an acceleration efficiency for
protons $\zeta_{\rmn{p}}\sim0.0001-0.01$, and for electrons
$\zeta_{\rmn{e}}\sim 0.001$.

{\bf Conclusions.}  The standard reacceleration model for RHs requires
a population of seed electrons to undergo turbulent
reacceleration. These seeds are generally thought to be secondary
electrons from hadronic CRp interactions. In this work we use
cosmological simulations to derive a population of seed CRps
originating from structure formation shocks and merger shocks during
the cluster build up. The resulting secondary population is
inconsistent with RH observations. We propose 3 possible solutions
that all produce gamma-ray emission below current upper limits and
that reproduce both the spectrum and the surface brightness profiles
of the Coma RH: (i) injected turbulence that is flatter than in
previously adopted models, (ii) streaming CRs, (iii) shock accelerated
CR electrons injected with $K_{\rmn{ep}} \sim 0.1$. We will pursue
further implications and distinguishing characteristics of these
competing models in future work.

{\bf Acknowledgments.} We thank Josh Wiener for discussions on CR
streaming. We are also grateful to Lawrence Rudnick for discussion on
uncertainties in the 1.4 GHz radio data. A.P. is grateful to the
Swedish research council for financial support. S.P.O. thanks NASA
grant NNX12AG73G for support. C.P.~gratefully acknowledges financial
support of the Klaus Tschira Foundation.
\vspace{-0.7cm}

\bibliography{bibtex/paper}

\end{document}